\documentclass[%
 reprint,
superscriptaddress,
 amsmath,amssymb,
 aps,
floatfix,
]{revtex4-2}

\usepackage{graphicx}% Include figure files
\usepackage{dcolumn}% Align table columns on decimal point
\usepackage{bm}% bold math
\usepackage[colorlinks,linkcolor=blue,anchorcolor=blue,urlcolor=blue, citecolor=blue]{hyperref}
\usepackage{float}
\usepackage{physics}
\usepackage{lipsum}  
\usepackage{mathtools}
\usepackage{subfig}
\usepackage{array}
\usepackage{diagbox}
\usepackage{caption}
\usepackage{siunitx}
\usepackage{placeins}
\usepackage{afterpage}
\usepackage[table]{xcolor}
\usepackage{etoolbox}
\usepackage{multirow}
\usepackage{xr}
\usepackage{subcaption}
\usepackage{bibunits}

\defaultbibliographystyle{apsrev4-2} % Keeps your standard APS formatting

\newcommand{\DTUAffiliations}{
\affiliation{Center for Macroscopic Quantum States (bigQ), Department of Physics, Technical University of Denmark, Kongens Lyngby, Denmark}
}

\begin{document}
\begin{bibunit} % <--- START MAIN TEXT BIBUNIT
\preprint{APS/123-QED}

\title{Practical continuous-variable quantum key distribution with squeezed light}% Force line breaks with \\
% \thanks{A footnote to the article title}%

\author{Huy Q. Nguyen}%
\email{hqung@dtu.dk}
\DTUAffiliations
\author{Ivan Derkach}
\affiliation{Department of Optics, Faculty of Science, Palacky University, Olomouc, Czech Republic}%
\author{Akash nag Oruganti}
\affiliation{Department of Optics, Faculty of Science, Palacky University, Olomouc, Czech Republic}%
\author{Adnan A.E. Hajomer}
\DTUAffiliations
\author{Hou-Man Chin}
\affiliation{Celare Quantum Communications ApS, Frederiksberg, Denmark}
\author{Radim Filip}
\affiliation{Department of Optics, Faculty of Science, Palacky University, Olomouc, Czech Republic}%
\author{Ulrik L. Andersen}
\DTUAffiliations
\author{Vladyslav C. Usenko}
\affiliation{Department of Optics, Faculty of Science, Palacky University, Olomouc, Czech Republic}%
\author{Tobias Gehring}
\DTUAffiliations
\affiliation{Celare Quantum Communications ApS, Frederiksberg, Denmark}

\date{\today}

\begin{abstract}
Continuous-variable quantum key distribution (CV-QKD) has gathered significant interest for its potential to achieve high secret key rates and seamless integration with existing optical communication infrastructure. State-of-the-art CV-QKD systems primarily use coherent states for simplicity. However, squeezed states of light have been theoretically shown to offer significant advantages, including higher secret key rates, greater resilience to excess noise, and reduced requirements on information reconciliation efficiency. In this work, we experimentally verify these theoretical predictions and propose and demonstrate a practical squeezed-state CV-QKD system based on modern local-local oscillator and digital-signal-processing techniques. Operating over fiber channels and considering finite-size security against collective attacks we show the advantages of our system over its coherent state counterpart. Our work paves the way for squeezed states to become practical resources for quantum key distribution and other quantum information protocols.
\end{abstract}

\maketitle

\section{Introduction}

Quantum key distribution (QKD) enables information-theoretic secure communication based on the principles of quantum physics \cite{pirandolaAdvances2020}. Among the different QKD flavors, continuous-variable quantum key distribution (CV-QKD) encodes information onto the quadratures of the optical field \cite{usenko2025continuousvariable}. Recent advances have established CV-QKD with practical composable security~\cite{jainPractical2022a}, enabling extended operational distances~\cite{hajomerLongdistance2024, zhangLongDistance2020} and increased secret key rates~\cite{hajomerContinuousVariable2024}.

Squeezed states form a foundational resource for numerous quantum information protocols. In quantum computation, they can be used for blind quantum computing \cite{tomoyukiContinuousVariable2012} and universal quantum computation with cluster states~\cite{MenicucciUniversal2006}, while recent experimental advances also demonstrate a quantum learning advantage~\cite{Zheng-HaoQuantum2025}. In quantum metrology, squeezed states have been shown to enable extreme sensitivity beyond the standard quantum limit and the ideal NOON state limit~\cite{nielsenDeterministic2023}, and to enable distributed quantum sensing with entangled probes~\cite{guoDistributed2020a}.

In the realm of quantum communication, squeezed states can be leveraged to implement oblivious transfer \cite{furrercontinuousvariable2018} and in CV-QKD to extend transmission distances~\cite{madsenContinuous2012}, enable advanced security \cite{gehringImplementation2015}, and eliminate information leakage in purely lossy channels~\cite{jacobsenComplete2018}. In quantum state discrimination, squeezed states can be used to surpass the coherent Helstrom bound~\cite{ChesiSqueezing-enhanced2018, walshallgaussianstate2025}, enabling advancements in quantum cryptography and quantum communication. Given their versatility and central role in these applications, squeezed states are a strong candidate for resource states in the emerging quantum Internet, enabling additional functionalities beyond QKD, especially since they are crucial for continuous variable (CV) entanglement swapping \cite{PolkinghorneContinuousVariable1999}, quantum teleportation \cite{FurusawaUnconditional1998}.

State-of-the-art CV-QKD systems typically utilize coherent signal states generated by laser sources, offering simplicity in implementation \cite{GG02,WeedbrookNoSwitch2004}. An alternative approach is to encode information onto quadrature-displaced squeezed states~\cite{CerfQuantumDistribution2001}, which was the initial proposal of Gaussian CV-QKD. The most inherent advantage of squeezed states is the increase in signal-to-noise ratio (SNR) due to the lower noise variance, which leads to higher mutual information. Furthermore, it has been shown that squeezed states minimize the information leakage and possibly completely remove the eavesdropper's information~\cite{jacobsenComplete2018,WinnelMinimization2021}. Interestingly, this make them extremely robust for free-space and satellite QKD~\cite{derkach2020squeezing,HosseinidehajSimple2022}. Theoretical studies have suggested that using squeezed states can allow CV-QKD systems to tolerate higher noise levels, to extend transmission distances, and to enhance the performance of systems with low information reconciliation efficiency~\cite{usenkoSqueezedstate2011, derkach2020squeezing, Oruganti_2025}.
 Despite these benefits, generating and detecting squeezed states presents additional complexity with previous experimental demonstration of squeezed states in CV-QKD~\cite{madsenContinuous2012,gehringImplementation2015,jacobsenComplete2018} were performed in laboratory conditions over short free-space channels and emulated the channel loss. Thus, they are quite far from full-scale implementation, enabling deployment in optical fiber infrastructure. An important step towards this goal is to demonstrate the advantage of QKD with squeezed states in a practical setting.

Significant progress has been made on developing compact squeezed sources~\cite{arnbakCompact2019} and implementing them on photonic integrated chips~\cite{zhangSqueezed2021, nehraFewcycle2022, parkSinglemode2024, ArgeDemonstration2025}. These developments paved the way for cheap and portable squeezed light sources. However, transmitting quadrature-squeezed states through an optical fiber and remotely detecting them with a homodyne detector remains challenging. Two major challenges to fiber-based squeezed-state CV-QKD are the phase shifts introduced by the fiber channel and the necessity for locally generated local oscillators (LLO) to avoid security loopholes~\cite{huangQuantum2013,jouguetPreventing2013}. Efforts to address these issues using optical phase-locked loops (OPLL)~\cite{suleiman40KmFiber2022} have encountered limitations, such as phase drift that destabilizes quadrature variances and correlations, complicating reliable key extraction~\cite{suleimanLongdistance2022}. Furthermore, an OPLL undoubtedly increases the system's overall complexity. Due to the above-described practical challenges, squeezed states were not, until now, considered to be a realistic competitor of coherent states in CV-QKD.

Over the last decade, DSP has served as a powerful workhorse for modern coherent-state CV-QKD systems~\cite{chinDigital2022,chinMachine2021,ZhangAutomatic2023,ErkilicEnhanced2025}. Leveraging techniques that have matured over more than half a century with digital computing and four decades alongside the Internet, DSP has significantly improved both the performance and ease of implementation of these systems. Recently, DSP has also been applied to enable the reconstruction of squeezed states and facilitate passive CV-QKD over deployed fiber~\cite{nguyenDigital2025}. Aside from the requirement of a squeezed-light source, this advancement allows squeezed-state CV-QKD to operate with simplified optical subsystems comparable to those of conventional coherent-state protocols.

In this work, we propose, experimentally demonstrate and theoretically analyze a prepare-and-measure (PM) CV-QKD protocol using quadrature-modulated squeezed states over optical fiber and effectively measured by radio frequency (RF) heterodyne detection. Through DSP, we eliminate the need for optical frequency and phase locking, thus enhancing the practicality of the protocol. We perform a security analysis taking into account finite-size effects and compare it with the conventional coherent states protocol under the same condition. Our results reveal a substantial enhancement in secret key rate with squeezed states compared to coherent states. Our analysis further shows that the squeezed-state CV-QKD protocol performs exceptionally well with lower reconciliation efficiency, reducing the computational demands of reverse reconciliation - a critical bottleneck in CV-QKD - thus promising a low-complexity, real-time and high-speed key generation system. Finally, we compare squeezed-state and coherent-state protocols under increasing excess noise levels, demonstrating that squeezed states are significantly more resilient. This resilience makes squeezed-state CV-QKD an attractive candidate for multiplexing with high-power classical channels in existing optical telecom infrastructure.

\section{Squeezed state CV-QKD protocol}

\subsection{Experimental implementation}
\begin{figure*}[ht]
    \centering
    \includegraphics[width=\linewidth]{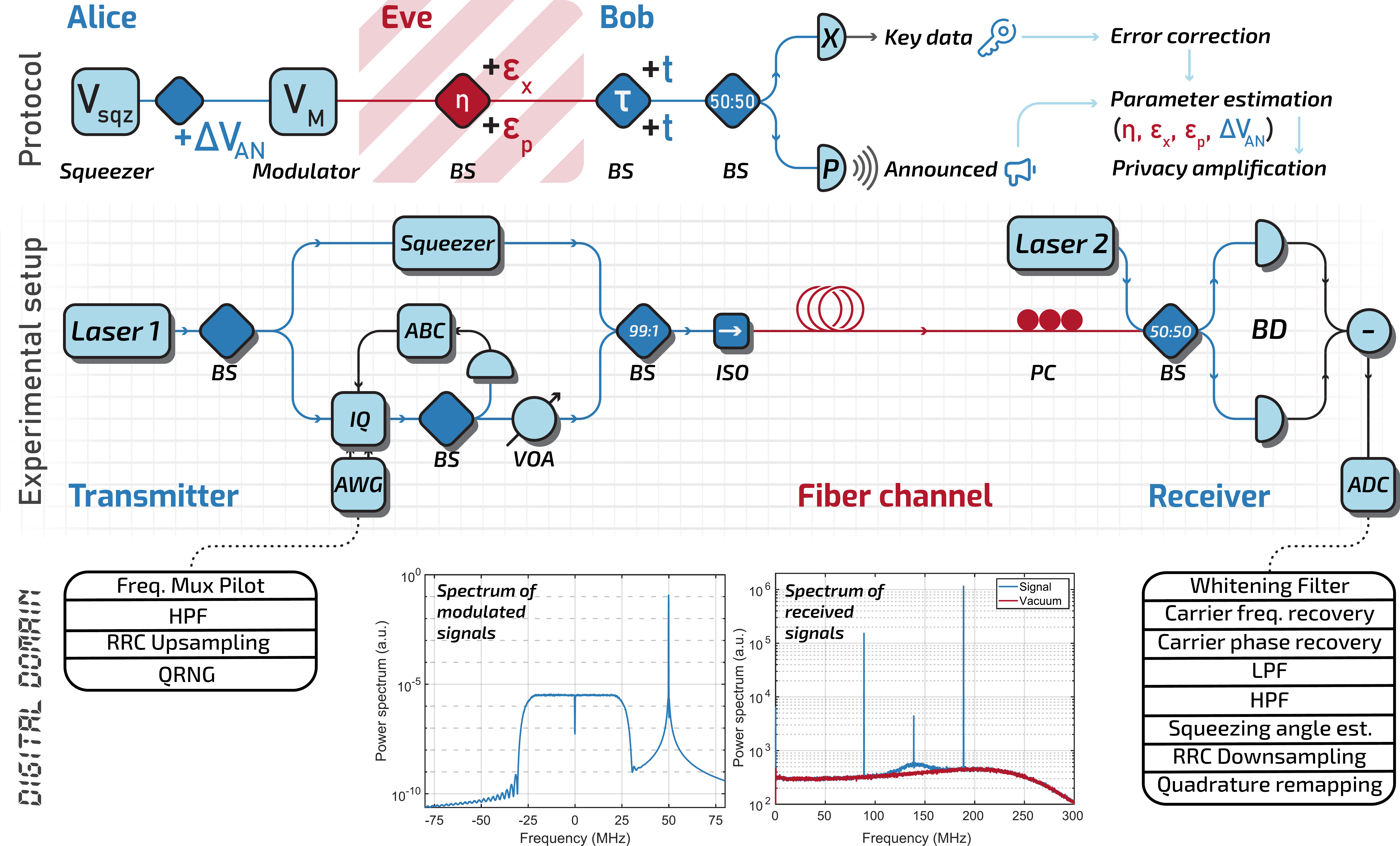}
    \caption{\textbf{Protocol}: Alice generates squeezed vacuum states with variance $V_{sqz}$ in x-quadrature, and $1/V_{sqz} +\Delta V_{AN}$ in p-quadrature, and modulates it in both quadratures (modulation variance $V_M$). Eve's channel subjects the signal to losses $\eta$ and quadrature dependent noise with variance $\epsilon_{x(p)}$. Bob performs heterodyne detection on the signal with an efficiency $\tau$ and added electronic noise $t$. The measurement outcomes of the p-quadrature are publicly announced and fully used for parameter estimation. The measurement outcomes of the x-quadratures are used to generate the keys. Here, we assume that we do error correction before parameter estimation \cite{LeverrierComposable2015,jainPractical2022a}. \textbf{Experimental setup}: IQ: in-phase and quadrature modulator, AWG: arbitrary waveform generator, ABC: automatic bias controller, VOA: variable optical attenuator, ISO: optical isolator, PC: polarization controller, BD: balanced detector, ADC: analog-to-digital converter, RRC: root-raised-cosine, QRNG: quantum random number generator, HPF: high-pass filter, LPF: low-pass filter.}
    \label{fig:sqz_qkd_setup}
\end{figure*}

The experimental setup is illustrated in Fig.~\ref{fig:sqz_qkd_setup}. 
On the transmitter side, squeezed light was generated at 1550 nm via spontaneous parametric down-conversion using a periodically poled potassium titanyl phosphate crystal in a hemilithic cavity. Details about the squeezed light source are available in Ref.~\cite{arnbakCompact2019}. Alice draws the displacement $\alpha$ of the quantum symbols with Gaussian statistics using a quantum random number generator (QRNG). The system operated at a rate of 50 MBaud. The quantum symbols were then upsampled using a root-raised cosine (RRC) pulse shape with a roll-off factor of 0.2. Baseband modulation was chosen to fully utilize the squeezing bandwidth. A 4th-order Butterworth high-pass filter (HPF) with a cutoff frequency of 170 kHz was applied to ensure the same temporal shaping as the mode detected by the receiver~\cite{hajomerModulation2022,chenContinuousmode2023}. A 50 MHz pilot tone was frequency-multiplexed to serve as a phase reference for the carrier recovery. The modulation waveform was used to drive an IQ modulator, and the optical signal strength was controlled with a VOA (variable optical attenuator).
The modulated signal was combined with the squeezed light at a 99:1 beam splitter, displacing the squeezed vacuum state in phase space with Gaussian statistics. An optical isolator was installed in the path to prevent back reflections from reaching the squeezed light source. The squeezed light source can be viewed as a modular add-on to a conventional coherent states system, enabling the implementation of squeezed states CV-QKD without major modifications. In fact, when benchmarking our proposed protocol against the coherent states protocol, the switch simply requires physically blocking the squeezed light from the free-space cavity.

At the receiver side, a separate free-running laser was used to generate a LLO. The quantum signal was measured using RF heterodyne detection with a balanced detector by frequency detuning the receiver laser with respect to the transmitter laser by an amount larger than the squeezing bandwidth~\cite{nguyenDigital2025}. After the measurement, we digitally reconstructed the displaced squeezed states as we describe in the following.

First, the frequency difference between the two lasers was estimated and corrected using the analytic representation of the pilot tone. After frequency offset compensation, we performed carrier phase recovery using an unscented Kalman filter (UKF) \cite{chinMachine2021}. To reduce out-of-band noise, particularly the high-power pilot tone, we applied a low-pass filter (LPF) around the quantum signal. The baseband modulation contained unwanted low-frequency components, such as the unsuppressed optical carrier and the automatic bias controller dither signal, which we eliminated using a HPF \cite{hajomerModulation2022}.
Following carrier phase recovery, we corrected the phase noise between the two lasers. However, the recovered quadratures did not necessarily align with the squeezing and anti-squeezing quadratures. Due to the intrinsic asymmetry of the squeezed state ensemble, any phase misalignment produces correlations between the X and P quadratures. By minimizing these correlations, we identified the optimal rotation needed to align Bob’s quadratures with the squeezing and anti-squeezing quadratures \cite{nguyenDigital2025}. However, because the squeezing field was not locked to the modulated coherent states, there is still a mismatch between Alice's modulation quadratures and Bob's measured quadratures.
In coherent-state systems, quadrature remapping has been used to rotate the received symbols, maximizing the covariance between Alice’s and Bob’s symbols \cite{qiGenerating2015}. We applied this technique to Alice’s symbols, recovering both the modulation angle and the covariance between Alice's and Bob's states. This method is applicable only under the conditions of symmetric modulation -- where the modulation variance is identical across both quadratures -- and equal detection efficiencies for both quadratures of the optical field.

\subsection{Security of the protocol} \label{sec:security}

Unlike coherent states, squeezed states are never perfectly pure in practice, posing a security risk as uncharacterized impurities may be exploited by an eavesdropper \cite{derkach2020squeezing, Oruganti_2025}. While impurity can be incorporated into the security analysis, it reduces performance and still relies on accurate squeezing level monitoring—risking underestimation of eavesdropper influence. Alternatively, continuous monitoring of both squeezing and purity allows treating the additional noise as trusted preparation noise. Since the most conservative security analysis assumes maximum added noise~\cite{Oruganti_2025}, all preparation noise can be attributed to the anti-squeezed quadrature, represented as $\Delta V_{AN}$.

Early theoretical works \cite{ralph1999continuous, cerf2001quantum, garcia2009continuous, usenkoSqueezedstate2011} on squeezed-state CV-QKD did not address signal-state verification in practical settings. Traditional protocols switch quadratures for squeezing and key encoding, with homodyne measurements contributing to either the raw key or state monitoring. Although this approach theoretically performs well and tolerates high noise \cite{usenkoSqueezedstate2011, madsenContinuous2012, derkach2020squeezing}, it requires careful resource allocation in the finite-size regime \cite{Oruganti_2025}. Measuring both quadratures simultaneously allows continuous key generation and real-time state verification \cite{garcia2009continuous}, enhancing security at the cost of reduced detection efficiency and added vacuum noise—a cost that can be partially mitigated by publicly disclosing measurements of the anti-squeezed quadrature. The security analysis of such protocol is based on the three-mode entanglement-based purification model detailed in Ref. \cite{usenkoSqueezedstate2011, Oruganti_2025} and Supplementary Material S.~\ref{supp:security-model}. In the asymptotic regime, the secret key fraction (SKF) secure against collective attacks is given by:
\begin{equation}\label{eq:SKR-base}
    K_{\infty}=\max \{0, \beta\,I(A,B) - \chi(B:E) \},
\end{equation}
    where $\beta$ is the reconciliation efficiency, $I(A, B)$ is the mutual information between Alice and Bob in the squeezed quadrature, and $\chi(B: E)$ is Holevo bound - the upper bound on the accessible information of Eve on homodyne measurement results used for the key generation. The measurements and public disclosure of the anti-squeezed quadrature aids parameter estimation (see Supplementary material S.~\ref{supp:security-model} and S.~\ref{supp:param_est}) and partially reduces the entropy of eavesdroppers state bounded by $\chi(B: E)$, which now becomes conditioned on the announcement of the non-signal quadrature $B^P$. The Holevo information $\chi(B:E)$ reads:
\begin{equation}
    \chi(B:E)=S(\rho_{AB^X|B^P})- S(\rho_{A|B^XB^P}).
    \label{eq:holevo}
\end{equation}
This approach has been shown to be more efficient than conventional homodyne-based protocol, especially in high-noise quantum channels, and for smaller block sizes ~\cite{Oruganti_2025}. For further details on the evaluation of respective elements of the key rate, (\ref{eq:SKR-base}) and purification model of the protocol see Ref.~\cite{Oruganti_2025}.

Usually, detector efficiency is not considered in parameter estimation. However, detector inefficiency reduces correlations between Alice's and Bob's data, which are essential for accurately estimating the channel transmission. This inefficiency consequently deteriorates the quality of the transmission estimate. Therefore, we extend the parameter estimation from~\cite{Oruganti_2025} to explicitly include detector efficiency.

The variance of the transmission estimator, taking detector efficiency into account, is given by:
\begin{equation}
\text{Var}[\hat{\eta}] =\sigma_\eta^2 \approx \frac{\eta \left[ V'_{Np} + 4 \eta V_M + V'_{Nx}\right]}{2 n V_M} 
\end{equation}

Here,
\begin{align*}
V'_{Nx} &= \frac{2}{\tau} + \eta(\epsilon + V_{\text{sqz}} - 1) + \frac{1-\tau}{\tau} (V_D-1), \\
V'_{Np} &= \frac{2}{\tau} + \eta(\epsilon + \frac{1}{V_{\text{sqz}}} + \Delta V_{AN} - 1) + \frac{1-\tau}{\tau} (V_D-1).
\end{align*}
The variance of the excess noise estimator in the X and P quaratures reads:
\begin{align}
    \text{Var}[\hat{\epsilon}_x] = \sigma_{\hat{\epsilon}_x}^2\approx \frac{2}{n}  [V'_{Nx}]^2 + (1-V_S)^2 \text{Var}[\hat{\eta}]\\
    \text{Var}[\hat{\epsilon}_p] =\sigma_{\hat{\epsilon}_p}^2\approx \frac{2}{n}  [V'_{Np}]^2 + (1-1/V_S - \Delta V_{AN})^2 \text{Var}[\hat{\eta}]
\end{align}

The secret key fraction taking into account finite-size effect reads:
\begin{align}   \label{eq:finite-key-1}
   K_{\text{finite}}  = & K_{\infty}(\eta_{\text{low}},\epsilon_{\text{up}})- \Delta(n) = \\ & \text{max}\{0, \beta I\left(\eta_{low},\epsilon_{up}\right) - \chi\left(\eta_{low},\epsilon_{up}\right) - \Delta(n)\}
   \label{eq:finite-size-key}
\end{align}
here, \(\Delta(n)\) is the asymptotic equipartition property correction term, which quantifies the finite-size penalty associated with replacing the smooth conditional min-entropy by its asymptotic conditional von Neumann entropy. This term vanishes in the limit \(n\to\infty\). The quantity \(n\) denotes the number of signals used for key extraction. Channel efficiency $\eta_{\text{low}}=\eta-6.5\sigma_{\hat{\eta}}$ and the excess noise $\epsilon_{\text{up}}=\epsilon + 6.5\sigma_{\hat{\epsilon}}$ are the worst case estimation of the corresponding to the estimation error probability of $10^{-10}$. Note that the excess noise here is the \textit{higher value} of the two quadratures. Details analysis of squeezed states CV-QKD taking into account finite-size effect under collective attacks can be found in Ref.~\cite{Oruganti_2025}. 

\section{Results and discussion}
\subsection{Squeezed states under imperfect reconciliation}

\begin{figure*}[ht]
    \centering
    \includegraphics[width=\linewidth]{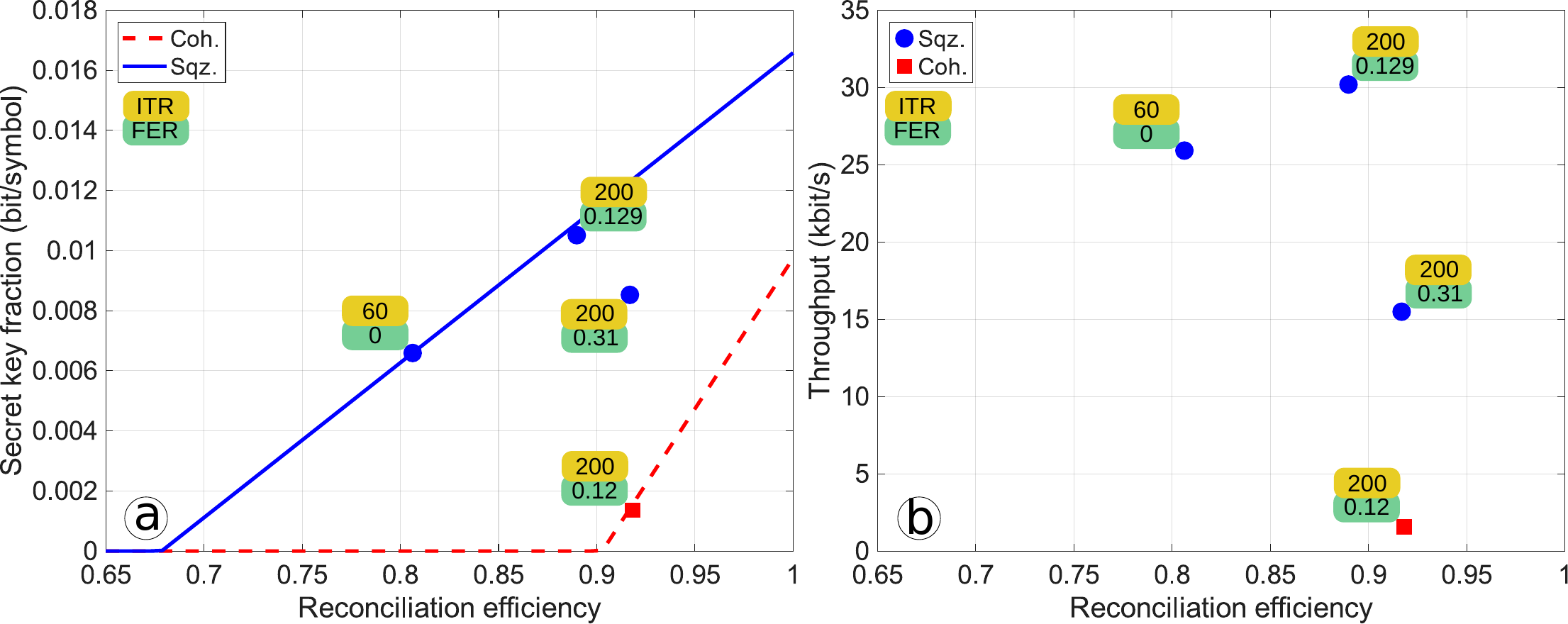}
    \caption{Key generation after 50 km fiber when varying reconciliation efficiency (a) Secret key fraction in bits per quantum symbol (b) Secret key throughput \textit{excluding DSP time}.}
    \label{fig:skf_vs_beta_throughput_vs_beta} 
\end{figure*}
\label{sec:sqz_imperfect_recon}
In CV-QKD, information reconciliation remains a critical bottleneck for real-time key generation. Since the signal-to-noise ratios (SNRs) are typically far below those of classical communication, error correction becomes a challenging task~\cite{maniMultiedgetype2021}. In practice, pushing the error correction code rate to approach the channel capacity yields high reconciliation efficiency but severely increases the frame error rate (FER) at low SNRs. As the reconciliation efficiency approaches unity, the code operates closer to the Shannon limit. This reduces the gap between the code rate and the channel capacity, leaving less tolerance for finite-size effects, decoder imperfections, and statistical fluctuations. As a result, the frame error rate may increase sharply in practical implementations. Moreover, maintaining this high efficiency requires hundreds of decoder iterations, which imposes a massive computational burden. Squeezed states provide an inherent advantage in this regard, as their lower noise variance leads to higher SNR and, in turn, higher mutual information (see Fig.~\ref{fig:mutualHolevo} in the Supplementary Material). A more detailed discussion of the information-theoretic reasons behind the robustness of squeezed states can be found in Supplementary Material S.~\ref{supp:physicalmeaning}.

To experimentally demonstrate the advantage of using squeezed states over coherent states in CV-QKD, we implemented both protocols over a 50-km ultra-low-loss (ULL) optical fiber with an total loss of of 7.25 dB. Experimental data was gathered from 250 frames, each comprising of $\approx4 \times 10^5$ quantum symbols resulting in $\approx10^8$ quantum symbols. The parameters estimated from the experiments are shown in Table~\ref{tab:parameters_estimation_50km}. The estimated squeezing variance was 0.417 SNU with the anti-squeezed noise $\Delta V_{AN} = 3.029$ SNU. The modulation variance was set to 1.372 SNU. Here, the excess noise $\epsilon$ is shown at the input of the channel, leading to $\epsilon_i = 2  u_i/\left(\eta \tau \right), \quad \text{for } i = x, p
$, where $u_i$ is the excess noise at detection and $\tau = 0.68$ is the trusted detection efficiency (See Supplementary Material S.~\ref{supp:param_est}).

\begin{table}[ht]
\centering
\begin{tabular}{|c|c|c|c|c|c|c|c|}
\hline
 & \begin{tabular}[c]{@{}c@{}}$V_{sqz}$\\ (SNU)\end{tabular} & \begin{tabular}[c]{@{}c@{}}$\Delta V_{AN}$\\ (SNU)\end{tabular} & \begin{tabular}[c]{@{}c@{}}$V_{M}$\\ (SNU)\end{tabular} & \begin{tabular}[c]{@{}c@{}}$\epsilon_x$\\ (SNU)\end{tabular} & \begin{tabular}[c]{@{}c@{}}$\epsilon_p$\\ (SNU)\end{tabular} & $\eta$ & \begin{tabular}[c]{@{}c@{}}$V_D$\\ (SNU)\end{tabular} \\ \hline
Sqz. & 0.417 & 3.029 & \multicolumn{1}{c|}{\multirow{2}{*}{1.372}} & 0.041 & 0.037 & 0.166 & 1.07 \\ \cline{1-3} \cline{5-8} 
Coh. &       &       &                                         & 0.032 & 0.031  & 0.163 & 1.07 \\ \hline
\end{tabular}
\caption{Parameters estimated from measurements with 50 km fiber channel. The parameters were estimated from $10^8$ quantum symbols.}
\label{tab:parameters_estimation_50km}
\end{table}

Error correction was performed with a multi-dimensional information reconciliation scheme implementing a multi-edge-type low-density-parity-check (MET-LDPC) error correction code \cite{maniMultiedgetype2021}. The base code rate of the original MET-LDPC code was 0.02, and puncturing was used to adjust the code rate to accommodate varying reconciliation efficiencies \cite{maniMultiedgetype2021}. Given a codeword length $n$, information bits $k$, and puncturing length $p$, the post-puncturing code rate is given by $R_\text{punc} = k/(n-p)$ \cite{maniMultiedgetype2021} which yields a code efficiency of $\beta = R_\text{punc} / I(A:B)$, where $I(A:B)$ is the mutual information (MI) of the classical information held by Alice and Bob after the quantum phase of the protocol. The parameters used for error correction are shown in Table~\ref{tab:coh_EC_param} and Table~\ref{tab:sqz_EC_param} in Supplementary Material S.~\ref{supp:info_recon_params} for the coherent states protocol and the squeezed-state protocol, respectively. 

When accounting for the frame error rate (FER) of the error correction process, the finite-size secret key fraction (Eq.~\ref{eq:finite-key-1}), expressed in bits per quantum symbol (bits/symbol), is given by:
\begin{equation}\label{eq:SKR}
   K_{\text{finite}} =
   (1-\text{FER})(R_\text{punc} - \chi(\eta_{\text{low}},\epsilon_{\text{up}}) - \Delta(n)).
\end{equation}

In this analysis, the MI was calculated based on the discretized quantum symbols held by Alice and Bob.

Fig.~\ref{fig:skf_vs_beta_throughput_vs_beta} (a) shows the results where the estimated SKF is plotted versus the error correction efficiency. The green bubbles represent the FER of the error correction process. The yellow bubbles indicate the minimal number of decoder iterations (ITR) for which we achieved the FER written in the green bubbles. Here, we did not aggressively minimize the number of decoder iterations as this would not make sense in a practical setting where the FER is also dependent on statistical fluctuations. Furthermore, the decoder implements early termination based on a syndrome check. As expected the number of required decoder iterations decreases with decreasing code efficiency as the code is further away from the Shannon capacity of the channel. The solid and dashed lines display the calculated SKF assuming an FER of zero. For high efficiencies, our experimental data points deviate from these lines as the FER becomes substantially larger than 0.

As seen in the figure, the performance degradation in the coherent state protocol is more pronounced as the $\beta$ decreases, while the squeezed-state protocol is less affected by this reduction. This experimentally demonstrates the theoretical prediction~\cite{usenkoSqueezedstate2011} that the secret key fraction scales differently with reconciliation efficiency for the two protocols. At a high reconciliation efficiency of $\beta\approx0.92$, the squeezed-state protocol achieves an SKF of 0.0085 bits/symbol, more than six times higher than the coherent states protocol, which yields 0.0014 bits/symbol. As $\beta$ is relaxed, both the FER and the required number of iterations decrease. This flexibility enables the squeezed-state protocol to achieve a peak SKF of 0.0105 bits/symbol at $\beta\approx0.89\%$ whereas the coherent states protocol cannot generate finite-size keys when $\beta < 0.9$.

The performance of CV-QKD systems is often limited by information reconciliation throughput. This throughput is inversely proportional to the number of decoder iterations. Given the lower efficiency of codes with fewer decoder iterations (Fig.~\ref{fig:skf_vs_beta_throughput_vs_beta} (a)), a new trade-off must be considered. We tested this by observing the throughput of an information reconciliation implementation on an NVIDIA GPU (2048 cores, 1.3 GHz), with results shown in Fig.~\ref{fig:skf_vs_beta_throughput_vs_beta} (b). Compared to Fig.~\ref{fig:skf_vs_beta_throughput_vs_beta}(a), we observed a shift in the optimal reconciliation efficiency ($\beta$) for finite-size throughput, which peaked at $\beta \approx 0.801$ for squeezed states. At this optimum, the advantage of squeezed states was evident, yielding a throughput of 30.19 kbit/s, nearly 20 times greater than the 1.57 kbit/s produced by the coherent state protocol. Note that given the same number of decoder iterations, the coherent state protocol's decoding time is about double that of the squeezed state protocol, as the latter generates a key from only the squeezed quadrature, while the former uses both.

\subsection{Squeezed states under high excess noise}
\label{sec:sqz_ex_noise}
Excess noise is another major bottleneck that directly degrades CV-QKD performance \cite{hajomerLongdistance2024}. An increase in noise leads to an increase in information leakage to the eavesdropper. Squeezed states have been shown to be very effective at minimizing Holevo information \cite{garcia2009continuous, jacobsenComplete2018}. Since squeezed states protocol only uses one quadrature to generate keys (compared to two for the coherent states), the question becomes how fast Holevo information grows with increasing excess noise. The key intuition behind the advantage of squeezed states under high excess noise is that their information leakage grows much more slowly than that of coherent states. A more detailed analysis of the information-theoretic basis behind the squeezed-state protocol's robustness against excess noise can be found in Supplementary Material S.~\ref{supp:physicalmeaning} (see Fig.~\ref{fig:bitsvsnoise}).

\begin{figure}[ht]
    \centering
    \includegraphics[width=\linewidth]{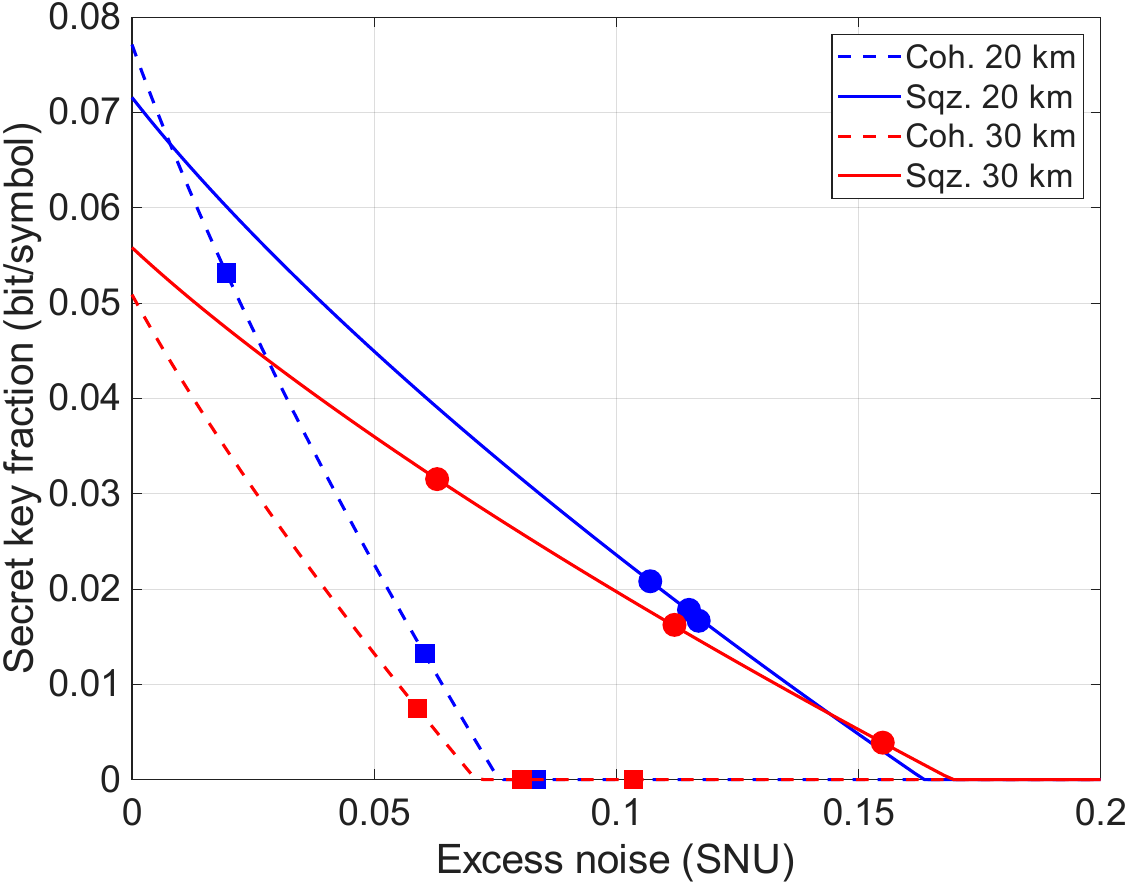}
    \caption{Secret key fraction (SKF) after 20 km and 30 km fiber at different levels of excess noise.}
    \label{fig:skf_vs_noise} 
\end{figure}

To demonstrate the noise tolerance advantage of squeezed states in CV-QKD, we introduced controlled noise during modulation and tested the system across two different channel lengths. Experimental data was gathered from 250 frames, each comprising of $\approx4 \times 10^5$ quantum symbols resulting in $\approx10^8$ quantum symbols. Theoretical simulations (solid and dashed lines in Fig.~\ref{fig:skf_vs_noise}) were conducted under conditions of fixed channel attenuation for each scenario. Here, we assumed $92\%$ reconciliation efficiency with $\text{FER=0}$. Table~\ref{tab:excess_noise_results_20km} and Table~\ref{tab:excess_noise_results_30km} in Supplementary Material S.~\ref{supp:ex_noise_params} present details on the parameter estimations and SKF for the squeezed and coherent state protocols at various excess noise levels. For the squeezed-state protocol, the noise floor without induced excess noise is estimated rather than directly measured, as explained in Supplementary Material~\ref{supp:param_est}, and the noise across both quadratures is conservatively symmetrized to the higher value. As a result the squeezed-state protocol generally shows higher intrinsic noise levels than the coherent states protocol.

Fig.~\ref{fig:skf_vs_noise} shows the experimental results and a fitted theory model. It is clearly visible that the two protocols have a different scaling in key rate when excess noise is increased. Starting at low excess noise levels we can see from the theory curves that the coherent states outperform squeezed states since squeezed states only utilize the squeezing quadrature to generate the keys. 
However, as noise increases, squeezed states begin to demonstrate superior performance. For instance, for the 20 km channel, at $\approx117$ mSNU of excess noise, the squeezed-state protocol achieves an SKF of 0.0167 bits/symbol, whereas the coherent states protocol, at a lower noise level of $\approx61$ mSNU, only achieves 0.00132 bits/symbol and fails to produce a finite-size positive secret key above $84$ mSNU of excess noise.

This trend is even more pronounced over the 30 km channel, where squeezed states consistently outperform coherent states across all experimental excess noise levels. At a lower excess noise level (just above 60 mSNU), the squeezed-state protocol achieves a finite-size key rate of 0.0315 bits/symbol, more than 4 times the rate of the coherent states protocol at 0.0074 bits/symbol. As excess noise increases, the coherent states protocol cannot tolerate around 80 mSNU, whereas the squeezed-state protocol can generate secret keys at noise levels exceeding 175 mSNU. 

From the theory curves we find that the QKD protocol operating over the 30 km channel tolerates more noise than over the 20 km channel. This is due to differences in both the anti-squeezed noise and the modulation variance between the two distances, with the 30 km channel's modulation variance being closer to its optimal value.

\section{Conclusion}

The benefits of using squeezed states in CV-QKD have been explored extensively in prior theoretical studies~\cite{usenkoSqueezedstate2011, Oruganti_2025}. While significant experiments have been conducted with channel loss simulated via optical attenuators, implementing CV-QKD systems using squeezed states over actual fiber channels remains challenging due to the complexity of generating and detecting squeezed states. In this work, we present the first practical demonstration of a CV-QKD system using squeezed states transmitted over fiber channels. By incorporating a specialized DSP chain, we significantly simplify the application of squeezed states in CV-QKD, preserving several key advantages over coherent-state protocols. 

First, our experimental results show that the squeezed-state protocol outperforms the coherent-state protocol, particularly under conditions of low reconciliation efficiency. 

To achieve high reconciliation efficiency the decoder requires many iterations, reducing its throughput. Squeezed states help alleviate the strain on classical post-processing, paving the way for high-speed, real-time key generation in CV-QKD systems.

Second, we demonstrated that squeezed light can improve CV-QKD systems' tolerance to excess noise. With QKD systems reaching closer to end-users, it becomes necessary to multiplex the quantum signal with strong classical light. CV-QKD has long been considered to be a good contender for this. With the ability to maintain high performance in noisy channels, squeezed light promises to be a valuable resource when integrating CV-QKD systems into existing optical telecom networks.

Recent efforts to increase the speed and practicality of CV-QKD systems entail the use of discretely modulated (DM) protocols. DM CV-QKD can reduce the stringent bit-resolution requirements of digital systems, thereby reducing implementation costs and increasing operating speeds while also reducing the complexity of classical post-processing~\cite{LiuHomodyne2021,hajomerContinuousVariable2023,ng2025gigabitratequantumkeydistribution}. Expanding the application of discrete modulation from coherent states to squeezed states offers a promising avenue to enhance systems performance and further exploit the inherent advantage of the non-classical states.  However, theoretical techniques need to be improved to expand to DM squeezed states especially in practical setting.

Beyond conventional fiber channels, free-space CV-QKD serves as a natural alternative when fiber-optic infrastructure is either unavailable or impossible to deploy securely~\cite{yinall-day2025,zhengfree-space2025}. Furthermore, satellite-based quantum communication offers an appealing route toward a global-scale, intercontinental quantum network~\cite{liaoSatellitetoground2017,limicrosatellitebased2025}. Theoretically, squeezed states are proven to significantly enhance the robustness of CV-QKD against high channel loss, modulation noise, and atmospheric turbulence~\cite{HosseinidehajSimple2022, derkach2020squeezing}. Although the distribution of squeezed light through atmospheric channels has been previously demonstrated, it relied on complex locking schemes~\cite{PeuntingerDistribution2014}. The simplicity and practicality of our proposed method make it directly applicable to free-space channels, thus opening a promising avenue for enhancing free-space CV-QKD performance.

The technique developed in this work could also benefit other quantum information and quantum metrology applications. Beyond QKD, quantum secret sharing (QSS) enables the secure distribution of information among multiple parties, provided a sufficient number of them remain honest~\cite{HilleryQuantum1999,CleveHowTo1999}. While QSS has been demonstrated using discrete variables (DV)~\cite{SchmidExperimental2005,ChenExperimental2005}, recent advancements include source-device-independent protocols that are robust against coherent attacks~\cite{XiaoExperimental2025}. In the CV domain, QSS has been realized with entangled states~\cite{LanceTripartite2004}, and more recently, through practical implementations using coherent states~\cite{LiuExperimental2023} and locally generated local oscillators~\cite{LiaoPractical2025}. The modulated squeezed states with techniques outlined in this paper can serve as a key resource to enhance the performance of these practical CV-QSS protocols.

Furthermore, squeezed states have long been shown to provide significant enhancements in metrology. For instance, the scheme in Ref.~\cite{guoDistributed2020a} demonstrated that displaced single-mode squeezed states can achieve distributed phase sensing beyond the standard quantum limit. However, that work was a proof-of-concept experiment involving free-space distribution on an optical table with a simulated phase shift. Our method extends this scheme to deployed fiber networks, thus significantly expanding the practicality and reach of the technology.

Due to implementation challenges, squeezed-state CV-QKD has not been a major focus of active research, with most previous work limited to theoretical analyses or proof-of-concept experiments. Our work demonstrates that modern techniques from coherent-state CV-QKD systems can be effectively extended to squeezed states. Combined with recent advances in photonic integration, our results position squeezed states as a strong candidate not only for CV-QKD but also for broader quantum information applications. These findings highlight the need for renewed focus and exploration in this promising area.

\section*{Acknowledgments}
We thank Naja Lautrup Nysom for help with the error correction framework.

We thank OFS optics for providing the SCUBA-125 fiber for this experiment.

\textbf{Funding:} The authors acknowledge support from the Danish Independent Research Council (grant agreement no. 0171-00055B), from the Carlsberg Foundation (project CF21-0466), from the QuantERA II Programme (project CVSTAR) funded by EU Horizon 2020 research and innovation program (grant agreement no. 101017733), from Innovation Fund Denmark (CyberQ, grant agreement no. 3200-00035B), and the DNRF Center for Macroscopic Quantum States (bigQ, DNRF142). This project has received funding from the European Union’s Horizon Europe research and innovation programme under the project ``Quantum Security Networks Partnership'' (QSNP, grant agreement No 101114043). I.\,D.\, acknowledges the project 22-28254O of the Czech Science Foundation. V.\,U.\,acknowledges the project 21-44815L of the Czech Science Foundation and the project CZ.02.01.01/00/22$\_$008/0004649 (QUEENTEC) of the Czech MEYS. The data was processed using the resources from DTU Computing Center \cite{HPC_cite}.

\putbib[apssamp]

\end{bibunit} % <--- END MAIN TEXT BIBUNIT
\clearpage
\onecolumngrid 

\begin{center}
    \vspace{0.1in}
    \textbf{\Large Supplementary Material:} \\
    \textbf{\large Practical continuous-variable quantum key distribution with squeezed light} \\
    \vspace{0.1in}
\end{center}
\vspace{0.2in}

% Switch back to two columns for the text
% \twocolumngrid 

\appendix
% Reset counters
\setcounter{equation}{0}
\setcounter{figure}{0}
\setcounter{table}{0}
\setcounter{page}{1}

% Add 'S' prefix to figure, table, and equation numbering
\renewcommand{\theequation}{S\arabic{equation}}
\renewcommand{\thefigure}{S\arabic{figure}}
\renewcommand{\thetable}{S\arabic{table}}

\renewcommand{\appendixname}{S.}
\renewcommand{\thesection}{\Roman{section}}

\makeatletter
\renewcommand{\@biblabel}[1]{[S#1]}
\renewcommand{\citenumfont}[1]{S#1}
\makeatother
\begin{bibunit} % <--- START SUPP TEXT BIBUNIT

\section{Security model}\label{supp:security-model}
 
\begin{figure}[ht]
    \centering
    \includegraphics[width=0.6\linewidth]{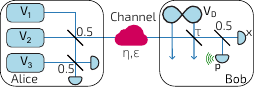}
    \caption{Equivalent entanglement based protocol}
    \label{fig:EB_model}
\end{figure}

The equivalent entanglement-based (EB) model is illustrated in Fig.~\ref{fig:EB_model}, where Alice holds the purification consisting of three single-mode squeezers, each with distinct squeezing parameters \cite{usenkoSqueezedstate2011,Oruganti_2025}. If Alice generates a pure single-mode squeezed states with the squeezing variance $V_S$ and displaces them with Gaussian statistics with a modulation variance $V_M$ in both quadratures, then the ensemble sent to Bob will have variances of $V_M + V_S$ in the $x$-quadrature and $V_M + 1/V_S$ in the $p$-quadrature. The variances of the three single-mode squeezers in Fig.~\ref{fig:EB_model} are sets as \cite{usenkoSqueezedstate2011}: 
\begin{equation}
\begin{split}
   & V_{1,2} = V_S + V_M \pm \sqrt{\frac{(V_S + V_M)\left(V_M+V_S V_M (V_S + V_M)\right)}{1 + V_S V_M}}\\
   & V_{3} = \frac{V_S^2 V_M (V_S + V_M)}{V_M (1 + V_S V_M)}
\end{split}
\end{equation}
If we can consider a noisy squeezed state with the variance in the squeezed quadrature $V_S$, the variance in the anti-squeezed will be $V_S + \Delta V_{AN}$. Here, $\Delta V_{AN}$ represents the anti-squeezed noise, which is assumed to be trusted and under Alice's control. For a detailed analysis of scenarios involving untrusted anti-squeezed noise, readers can refer to Ref.~\cite{Oruganti_2025}. When modulating these noisy squeezed states, the variance of the states coming out of the channel will be $V_M + V_S$ and $V_M + 1/V_S + \Delta V_{AN}$. This added noise was assumed to be trusted and was modeled as extra modulation in the anti-squeezed quadrature~\cite{Oruganti_2025}.

The thermal channel, presumed to be under Eve's control, is characterized by its efficiency $\eta$ and excess noise $\epsilon$. Imperfect detection is modeled by overlapping the mode with another TMSV with a variance $V_D$ on a beam splitter with the transmittance of $\tau$ after the channel. Bob then performs a heterodyne measurement on the received mode. Alice and Bob do not use the anti-squeezed quadrature for key generation but measurements in this quadrature are important for parameter estimation and system performance monitoring. In our protocol, Bob performs balanced heterodyne detection on the received mode and publicly announces the results of the anti-squeezing measurements (see Section II.B in the main text or Ref.~\cite{Oruganti_2025} for further details). 

\section{Parameter estimation}
\label{supp:param_est}
Parameter estimation is a crucial part of CV-QKD which allows the trusted parties to evaluate the information leakage from the untrusted channel. The channel is characterized by two parameters: the channel transmittance and the excess noise. After detection, the excess noise in conventional coherent states systems can be calculated with the conditional variance by
\begin{equation}
    u = V_{B} - C^{2}_{AB} - t - 1\ ,
    \label{eq:coh_excess_noise}
\end{equation}
where $V_{B}$ is the variance of the received signal from Bob normalized to shot noise, $C_{AB}^2$ is the normalized covariance between Alice and Bob at the receiver side, $t$ is the trusted electronic noise of the detector and the shot noise, or vacuum state noise variance is normalized to 1. However, in the case of modulated squeezed state QKD the beautiful ``1'' in Eq.~\ref{eq:coh_excess_noise} from the vacuum states is replaced by the respective variances of the squeezed and anti-squeezed noise of the squeezed state scaled by the transmittance of the channel. Since the channel is fully under Eve's control, this noise floor cannot be directly measured as this would lead to a security loophole where the eavesdropper can change it, leading to a misestimation of the excess noise. To circumvent this, our protocol employs a procedure to estimate the noise floor based on the squeezing and anti-squeezing levels at the beginning of the channel established through B2B measurement along with the estimated channel efficiency. 

In the B2B measurement, we measure a noisy squeezed state with an imperfect detector. Without loss of generality, we assume in the following that the generated states are squeezed in the $X$ and anti-squeezed in the $P$ quadrature. The variances measured in both quadratures read:
\begin{equation}
\begin{split}
    V_{X}^\text{b2b} = V_\text{sqz}^\text{b2b} + t \\
    V_{P}^\text{b2b} = V_\text{anti-sqz}^\text{b2b} + \frac{\tau}{2} \Delta V_{AN} + t \\
\end{split}
\end{equation}
where $V_{sqz}^{b2b}$ is the measured squeezing variance, $t$ is the electronic noise, and $\tau$ is the detection efficiency. The factor of $1/2$ comes from the heterodyne detection. Here, the noisy squeezed states were modeled by the pure squeezed states ($V^{pure}_{anti-sqz}=1/V^{}_{sqz}$) with added the anti-squeezed noise $\Delta V_{AN}$. The pure squeezing can be calculated by:
\begin{equation}
    V_{sqz}^\text{pure} = 1 - \left[ 1 - (V_X^\text{b2b} - t)\right]/\left(\frac{\tau}{2}\right )
\end{equation}

In the anti-squeezed quadrature, we separate the anti-squeeze variance into the contribution from the pure anti-squeezed and the anti-squeezed noise $ \Delta V_{AN}$:
\begin{equation}
\begin{split}
    V_{anti-sqz}^\text{b2b} = 1 +\frac{\tau}{2}\left(V_\text{sqz}^{-1} -1\right) \\
    \Delta V_{AN} = (V_P^\text{b2b} - V_\text{anti-sqz}^\text{b2b} - t)/\left(\frac{\tau}{2}\right )
\end{split}
\end{equation}
The PM covariance matrix after detection reads:
\begin{equation}
    \gamma_{PM}^{RX} = \begin{bmatrix}
        V_M     & 0 & \sqrt{\frac{\tau\eta}{2}}V_M &0 \\
        0 & V_M & 0 & \sqrt{\frac{\tau\eta}{2}}V_M \\
        \sqrt{\frac{\tau\eta}{2}}V_M & 0 & V_\text{Rx}^{X} & 0 \\
        0 & \sqrt{\frac{\tau\eta}{2}}V_M & 0 &  V_\text{Rx}^{P} \\
    \end{bmatrix}
\end{equation}
where:
\begin{equation}
    \begin{split}
     V_\text{Rx}^{X} &=  \frac{\tau\eta}{2} V_M  + u_x + t + V_\text{sqz}^\text{Rx}\\
     V_\text{Rx}^{P} &=  \frac{\tau\eta}{2} V_M + u_p + t + V_\text{anti-sqz}^\text{Rx} + \frac{\tau\eta}{2} \Delta V_{AN}
     \end{split}
\end{equation}
with $\eta$ being the channel transmission efficiency, $V^\text{Rx}_\text{sqz}$ and $V^\text{Rx}_\text{anti-sqz}$ the squeezing and anti-squeezing variance after detection which can be calculated via:
\begin{equation}
\begin{split}
V^\text{Rx}_\text{sqz} = 1 - \frac{\tau\eta}{2}\left(1 - V_\text{sqz}^\text{pure}\right) \\
V^\text{Rx}_\text{anti-sqz} = 1 + \frac{\tau\eta}{2}\left(V_\text{anti-sqz}^\text{pure} - 1\right) \\
\end{split}
\end{equation}
Note that instead of the noise floor of 1, right now the noise floor in the X and P quadrature are $V_\text{sqz}^\text{Rx}$ and $V_\text{anti-sqz}^\text{Rx} + \frac{\tau\eta}{2} \Delta V_{AN}$, respectively. Thus, the excess noise after detection can be estimated as:
\begin{equation}
\begin{split}
    u_X &= V_{X}^\text{det} - \frac{\tau\eta}{2} V_M - V^\text{Rx}_\text{sqz} - t \\
               &=  V_{X}^\text{det} - C_{AB}^2 - V^\text{Rx}_\text{sqz} - t \\
    u_P &= V_{P}^\text{det} - \frac{\tau\eta}{2} V_M - V^\text{Rx}_\text{anti-sqz} -\frac{\tau\eta}{2} \Delta V_{AN} - t \\
            &= V_{P}^\text{det} - C_{AB}^2- V^\text{Rx}_\text{anti-sqz} -\frac{\tau\eta}{2} \Delta V_{AN} - t
\end{split}
\end{equation}

% \section{Finite-size calculation}\label{appendix:finite-size-cal}

\section{Parameters for information reconciliation}
\label{supp:info_recon_params}
For the information reconciliation, we employ a MET-LDPC code with a code rate of 0.02 and the codeword length $n=819200$~\cite{maniMultiedgetype2021}. To investigate the performance of the CV-QKD protocols across different $\beta$, we adopted a rate-adaptive reconciliation scheme. 
We used the puncturing technique to increase the code rate~\cite{WangEfficientRate2017}. The puncturing length $p$ and the reconciliation efficiency $\beta$ for the coherent states protocol and the squeezed-state protocol are detailed in Table~\ref{tab:coh_EC_param} and Table \ref{tab:sqz_EC_param} respectively. 
\begin{table}[h!]
\centering
\begin{tabular}{|c|c|c|c|c|}
\hline
\begin{tabular}[c]{@{}c@{}}MI\\ (bits/quadrature) \end{tabular}       & R & Code length & p & $\beta$         \\ \hline
0.0502 & 0.02      & 819200      & 463483       & 0.918 \\ \hline
\end{tabular}

\caption{Parameters used for error correction in coherent-state protocol.}
\label{tab:coh_EC_param}
\end{table}

\begin{table}[h!]
\centering
\begin{tabular}{|c|c|c|c|c|}
\hline
\begin{tabular}[c]{@{}c@{}}MI\\ (bits/quadrature) \end{tabular}       & R & Code length & p & $\beta$        \\ \hline
0.0516   & 0.02      & 819200      & 472700       & 0.917 \\ \hline
0.0516   & 0.02      & 819200      & 462162       & 0.890 \\ \hline
0.0516   & 0.02      & 819200      & 425162      & 0.806 \\ \hline
\end{tabular} 
\caption{Parameters used for error correction in squeeze-state protocol.}
\label{tab:sqz_EC_param}
\end{table}

\section{Parameters for excess noise tolerance investigation}
\label{supp:ex_noise_params}
Parameter estimation and experimental results for the investigation of excess noise tolerance for coherent states and squeezed-state protocol. $V_\text{sqz}$ is the estimated squeezing variance before detection, $\Delta V_\text{AN}$ is the anti-squeezed noise variance, $V_\text{mod}$ is the modulation variance, $\xi_x$ and $\xi_p$ are the excess noise in the two quadrature referred to channel input, $\eta$ is the channel efficiency, $t$ is the trusted detection noise referred to channel output. All the variances are calculated in SNU. The figure-of-merit: the SKF is calculated in bits/symbol. Table \ref{tab:excess_noise_results_20km} and Table \ref{tab:excess_noise_results_30km} showed the estimated parameters and the calculated SKF after 20 km and 30 km respectively. 

\begin{table}[ht]
\centering
\begin{tabular}{|c|c|c|c|c|c|c|c|c|}
\hline
       &  \begin{tabular}[c]{@{}c@{}}$V_\text{sqz}$\\ {\scriptsize (SNU)} \end{tabular}  & \begin{tabular}[c]{@{}c@{}}$\Delta V_\text{AN}$\\ {\scriptsize (SNU)} \end{tabular}   & \begin{tabular}[c]{@{}c@{}}$V_{M}$\\ {\scriptsize (SNU)} \end{tabular}  & \begin{tabular}[c]{@{}c@{}}$\epsilon_x$\\ {\scriptsize (SNU)} \end{tabular} & \begin{tabular}[c]{@{}c@{}}$\epsilon_p$\\ {\scriptsize (SNU)} \end{tabular} & $\eta$ & \begin{tabular}[c]{@{}c@{}}$V_D$\\ {\scriptsize (SNU)}\end{tabular} & \begin{tabular}[c]{@{}c@{}}SKF\\ {}\end{tabular}\\  \hline
\multirow{3}{*}{Coh.} & \multirow{3}{*}{}       & \multirow{3}{*}{}       & \multirow{6}{*}{1.067} & 0.0195  & 0.0195  & 0.407  & 1.0731 & 0.0531
                   \\ \cline{5-9} 
                     &                         &                         &                         & 0.062  & 0.059  & 0.408 & 1.07 & 0.0132                   \\ \cline{5-9} 
                     &                         &                         &                         & 0.084  & 0.083  & 0.416 & 1.07 & blue{0}                    \\ \cline{1-3} \cline{5-9} 
\multirow{3}{*}{Sqz.} & \multirow{3}{*}{0.427} & \multirow{3}{*}{3.119} &                         & 0.072  & 0.107  & 0.413 & 1.07 &  0.0208                    \\ \cline{5-9} 
                     &                         &                         &                         & 0.117   & 0.092   & 0.410 & 1.07 & 0.0167                  \\ \cline{5-9} 
                     &                         &                         &                         & 0.115   & 0.106   & 0.414 & 1.07 & 0.0178                   \\ \hline
\end{tabular}
\caption{Parameter estimation and experimental results for the 20 km channel. The parameters were estimated from $10^8$ quantum symbols.}
\label{tab:excess_noise_results_20km}
\end{table}
\begin{table}[ht]
\centering
\begin{tabular}{|c|c|c|c|c|c|c|c|c|}
\hline
          &  \begin{tabular}[c]{@{}c@{}}$V_\text{sqz}$\\ {\scriptsize (SNU)} \end{tabular}  & \begin{tabular}[c]{@{}c@{}}$\Delta V_\text{AN}$\\ {\scriptsize (SNU)} \end{tabular}   & \begin{tabular}[c]{@{}c@{}}$V_{M}$\\ {\scriptsize (SNU)} \end{tabular}  & \begin{tabular}[c]{@{}c@{}}$\epsilon_x$\\ {\scriptsize (SNU)} \end{tabular} & \begin{tabular}[c]{@{}c@{}}$\epsilon_p$\\ {\scriptsize (SNU)} \end{tabular} & $\eta$ & \begin{tabular}[c]{@{}c@{}}$V_D$\\ {\scriptsize (SNU)}\end{tabular} & \begin{tabular}[c]{@{}c@{}}SKF\\ {}\end{tabular}\\  \hline
\multirow{3}{*}{Coh.} & \multirow{3}{*}{}       & \multirow{3}{*}{}       & \multirow{6}{*}{1.461} & 0.057  & 0.061  & 0.276 & 1.07 &  0.0074                    \\ \cline{5-9} 
                      &                         &                         &                         & 0.080  & 0.081  & 0.283 & 1.07 & 0                    \\ \cline{5-9} 
                      &                         &                         &                         & 0.104  & 0.103  & 0.274 & 1.07 & 0                       \\ \cline{1-3} \cline{5-9} 
\multirow{3}{*}{Sqz.} & \multirow{3}{*}{0.416} & \multirow{3}{*}{2.714} &                         & 0.017 & 0.063  & 0.291 & 1.07 & 0.0315                   \\ \cline{5-9} 
                      &                         &                         &                         & 0.081 & 0.112  & 0.292 & 1.07 & 0.0162                  \\ \cline{5-9} 
                      &                         &                         &                         & 0.103  & 0.155  & 0.291 & 1.07 & 0.0039                   \\ \hline
\end{tabular}
\caption{Parameter estimation and experimental results for the 30 km channel. The parameters were estimated from $10^8$ quantum symbols.}
\label{tab:excess_noise_results_30km}
\end{table}

\section{Information-theoretic origin of squeezed states protocol advantage}\label{supp:physicalmeaning}
\begin{figure*}[ht!]
    \centering
    \subfloat[\label{fig:bitsvsAtten}]{\includegraphics[width=0.32\linewidth]{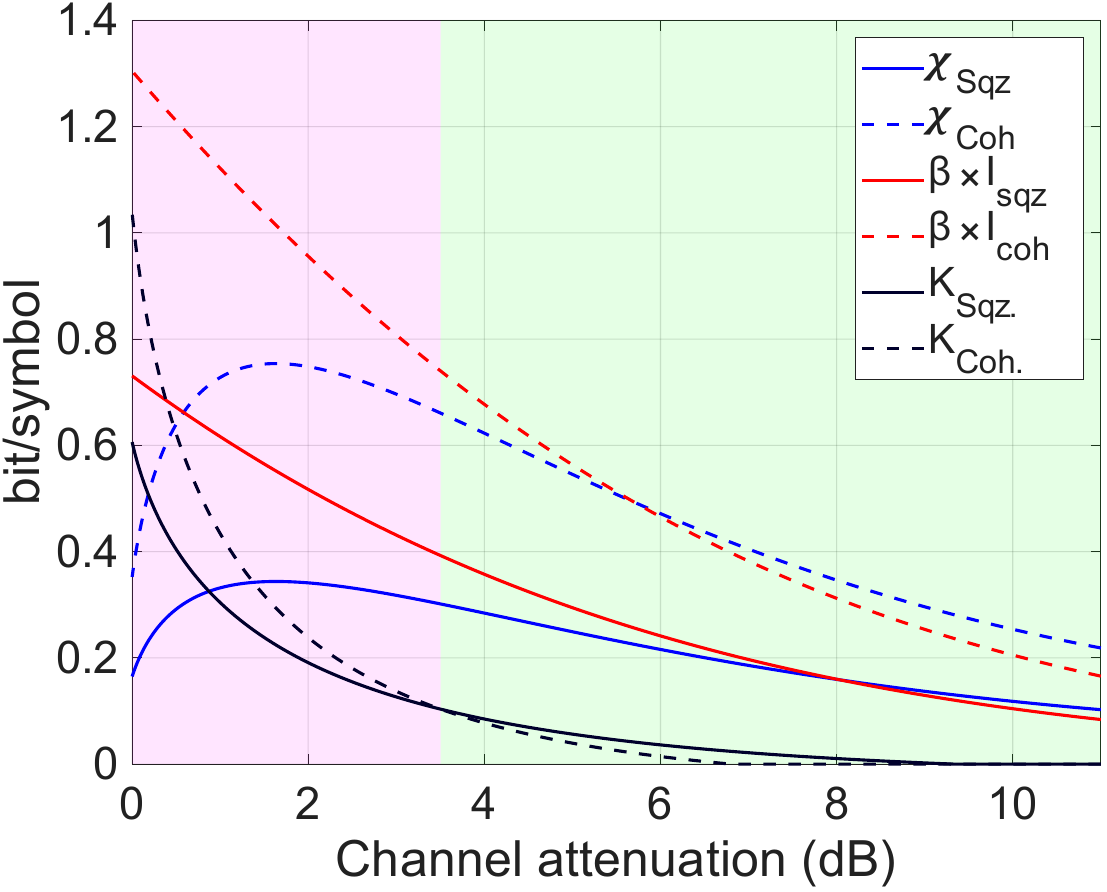}}
    \hfill
    \subfloat[\label{fig:bitsvsbeta}]{\includegraphics[width=0.32\linewidth]{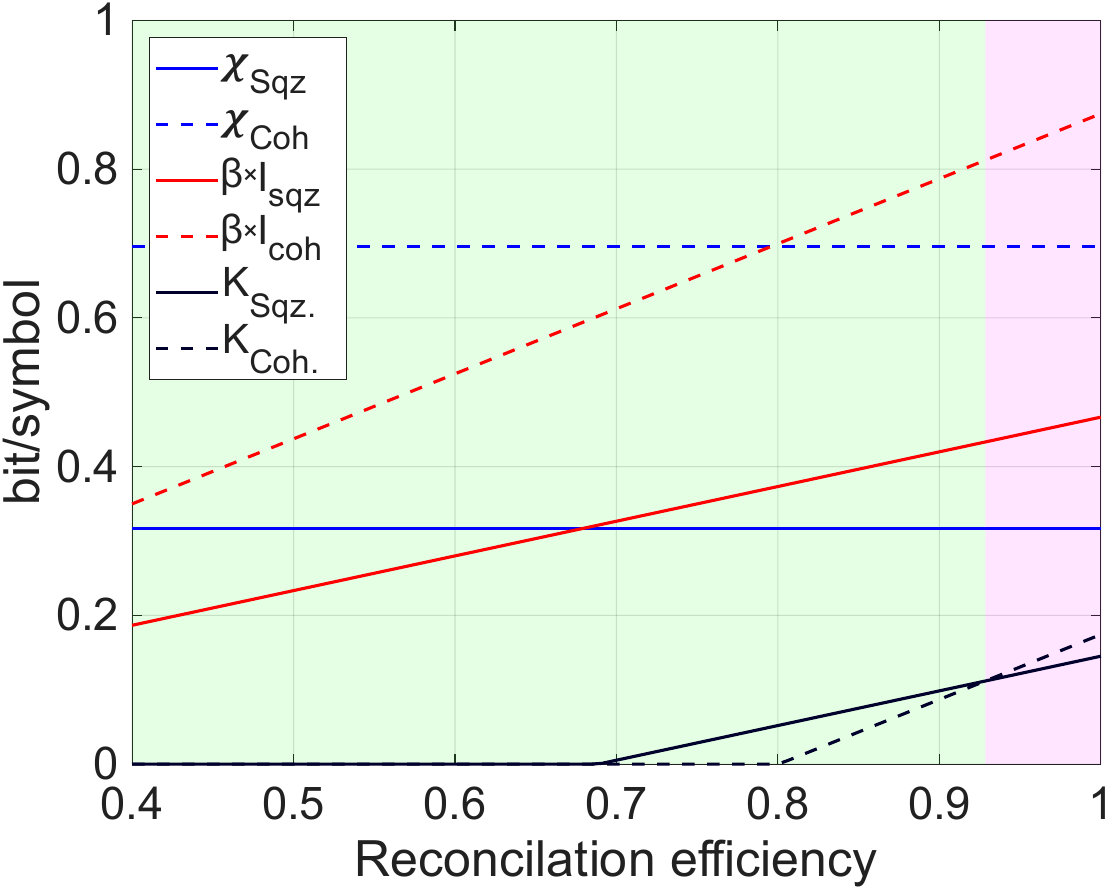}}
    \hfill
    \subfloat[\label{fig:bitsvsnoise}]{\includegraphics[width=0.32\linewidth]{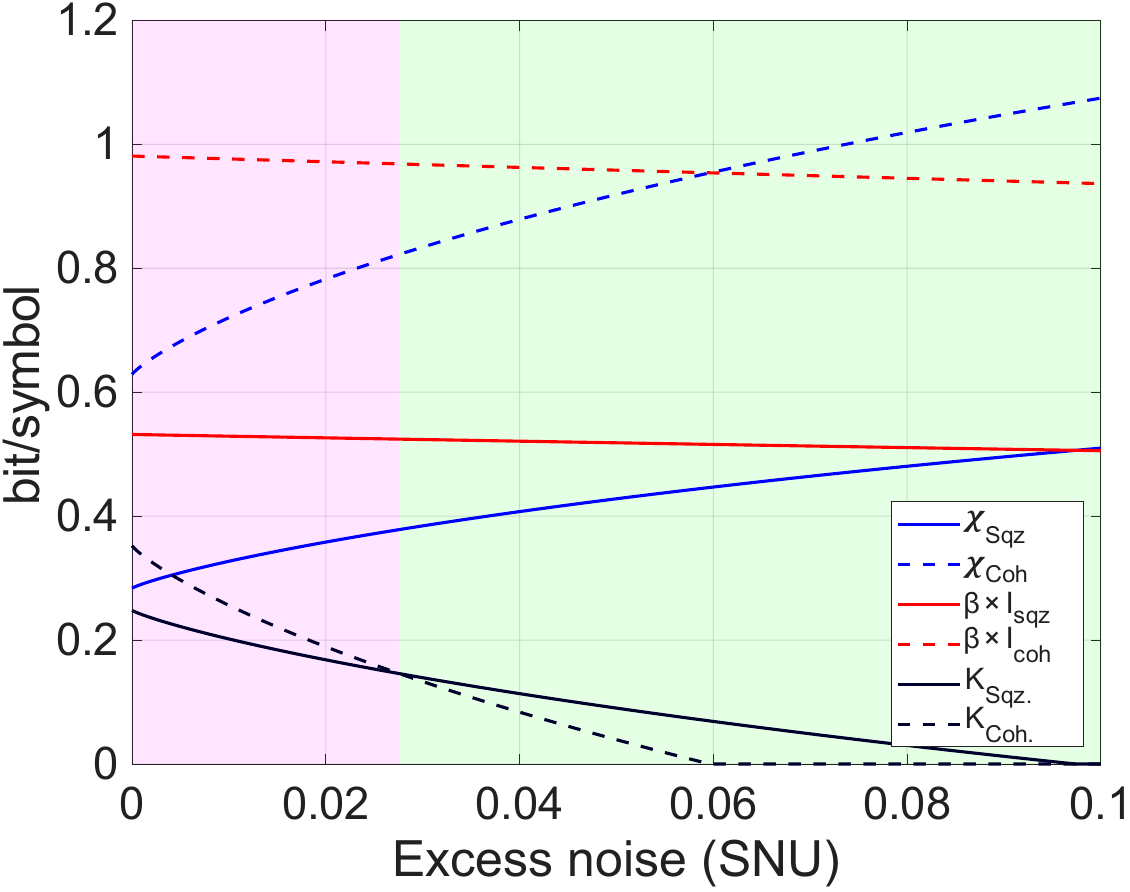}}
    
    \caption{Theoretical calculation of mutual information ($I_{sqz}$ and $I_{coh}$), Holevo information ($\chi_{sqz}$ and $\chi_{coh}$) and secret key rate ($K_{sqz}$ and $K_{coh}$) in asymptotic regime, when varying (a) channel attenuation, (b) reconciliation efficiency and (c) excess noise. The green regions show where squeezed states outperform coherent states protocol.}
    \label{fig:mutualHolevo}
\end{figure*}
When considering cases where \textit{both} coherent states and squeezed states protocols utilize a \textit{single quadrature} for key generation, the advantage of squeezed states can be readily explained by the increase in mutual information and the reduction of Holevo information. Previous studies have thoroughly investigated scenarios involving homodyne detection and uni-dimensional protocols \cite{usenkoSqueezedstate2011,jacobsenComplete2018,derkach2020squeezing}. In particular, Ref.~\cite{jacobsenComplete2018} demonstrated that the intuitive advantage of squeezed states lies in the reduction of Holevo information, which can even be completely eliminated in the case of a purely lossy channel. However, the requirement for homodyne detection at the channel output increases implementation complexity and limits the protocol's practicality. However, in this study, we compare two protocols: a conventional \textit{coherent states} protocol where secret keys are generated using \textit{both quadratures}, and a \textit{squeezed-state protocol} with heterodyne detection where secret keys are generated using \textit{only the squeezed quadrature}. Due to these differing configurations, explaining the inherent advantages of squeezed states becomes more nuanced. The secret key rate is determined by two nonlinear quantities: mutual information and Holevo information (Figure~\ref{fig:mutualHolevo}). It is the interplay between these two functions and the reconciliation efficiency that provides the advantage of squeezed states.

While the coherent states protocol yields higher mutual information by generating keys in both quadratures (providing almost twice the information), the Holevo information scales highly nonlinearly with variations in channel efficiency and excess noise (Figures~\ref{fig:bitsvsAtten} and~\ref{fig:bitsvsnoise}). When the reconciliation efficiency decreases, the Holevo information remains unaffected, but the effective mutual information between Alice and Bob decreases (Figure~\ref{fig:bitsvsbeta}). Here, due to the better signal-to-noise ratio provided by the squeezed states, the mutual information decreases more slowly than in the coherent states case. Conversely, an increase in excess noise reduces the mutual information similarly in both protocols, but the Holevo information increases much more drastically in the coherent states case. Ultimately, these factors lead to a cut-off point where squeezed states outperform coherent states (Figures~\ref{fig:bitsvsbeta} and~\ref{fig:bitsvsnoise}).

In fact, early developments in CV-QKD has considered measuring both quadratures of the squeezed signal state \cite{garcia2009continuous}, and have noted increased key rate in comparison with the coherent-state protocol. Although, the protocol used in our work differs (since we perform announcement of the non-signal quadrature) the rationale remains the same - heterodyne measurements introduce additional (trusted vacuum) noise that effectively counteracts the excess noise added in the channel. Latter is also known as "fighting noise with noise" \cite{usenko2016trusted}. Interestingly, the volume of such noise can be optimized, which is considered in  previous theoretical works \cite{garcia2009continuous, Oruganti_2025}.

\section{Squeezed states versus coherent states CV-QKD over long distance channel}\label{supp:longdistcomp}
\begin{figure*}[ht]
    \centering
        \includegraphics[width=\linewidth]{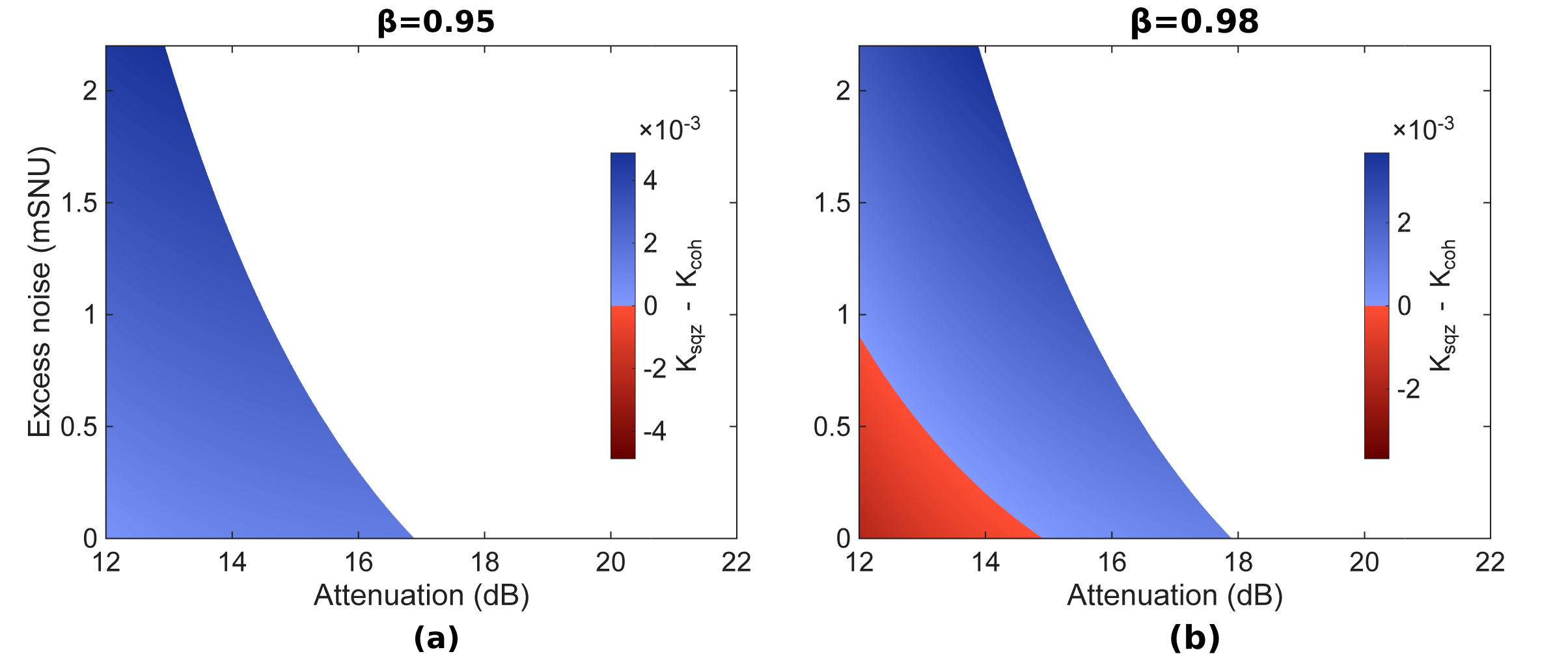}
\caption{Difference in the secure key rate between squeezed-state and coherent-state protocols ($K_{qz}-K_{coh}$, bit/symbol). The blue region indicates regimes where squeezed states outperform coherent states, the red region denotes where coherent states are superior, and the white region corresponds to parameters where neither protocol can generate a secure key. Excess noise is considered at the \textit{end of the channel}.}
\label{fig:key_rate_long_distance}
\end{figure*}
Figure~\ref{fig:key_rate_long_distance} shows the finite-size key rate difference between squeezed states and coherent states ($K_{qz}-K_{coh}$). The x-axis shows the channel attenuation, corresponding to approximately 80 km to 150 km of ultra-low-loss fiber, while the y-axis shows the excess noise at the channel output. In both protocols, the modulation variance was set to $V_{mod} = 5$ SNU, the trusted detection efficiency was $\tau=0.68$, the trusted noise was $V_D=0.07$ and the block size is $10^9$. The squeezing level was set to a realistic 3 dB with 3 SNU of anti-squeezed noise. Squeezed states continue to show superior performance even at longer distances. With a reconciliation efficiency of $\beta=0.95$, the squeezed-state protocol always outperforms coherent states (Figure~\ref{fig:key_rate_long_distance}a. Assuming a highly efficient reconciliation process ($\beta=0.98$), the coherent-state protocol yields a higher finite-size key rate only under conditions of low excess noise ( $< 1$ mSNU at the channel output) and channel attenuations below 15 dB (Bottom left part of Figure~\ref{fig:key_rate_long_distance}b. Thus, this analysis demonstrates that squeezed states not only maintain their performance advantage over long distance channels but also serve as a vital resource for extending the maximum transmission reach of CV-QKD systems.

\vspace{0.2in}
\begin{center}
    \textbf{\normalsize Supplementary References}
\end{center}
\putbib[apssamp]
\end{bibunit} 
\end{document}